\documentclass[aps,twocolumn,showpacs,preprintnumbers,prb]{revtex4}

\usepackage{graphicx}
\usepackage{dcolumn}
\usepackage{bm}

\begin{document}

\title{Self-energy and lifetime of Shockley and image states on Cu(100) and
Cu(111): Beyond the $GW$ approximation of many-body theory}

\author{M. G. Vergniory$^{1,2}$, J. M. Pitarke$^{1,3}$, and P. M. 
Echenique$^{2,4}$} 
\affiliation{
$^1$Materia Kondentsatuaren Fisika Saila, UPV/EHU, and Unidad F\'\i
sica Materiales CSIC-UPV/EHU,\\
644 Posta kutxatila, E-48080 Bilbo, Basque Country, Spain\\
$^2$Donostia International Physics Center (DIPC),\\
Manuel de Lardizabal Pasealekua, E-20018 Donostia, Basque Country, Spain\\
$^3$CIC nanoGUNE Consolider, Mikeletegi Pasealekua 56, E-2009 Donostia, Basque
Country, Spain\\
$^4$Materialen Fisika Saila, UPV/EHU, and Unidad F\'\i sica Materiales
CSIC-UPV/EHU,\\
1071 Posta kutxatila, E-20018 Donostia, Basque Country, Spain}

\date{\today}

\begin{abstract}
We report many-body calculations of the self-energy and lifetime of 
Shockley and image states on the (100) and (111) surfaces of Cu that go 
beyond the $GW$ approximation of many-body theory. The self-energy is 
computed in the framework of the $GW\Gamma$ approximation by including 
short-range exchange-correlation (XC) effects both in the screened 
interaction W (beyond the random-phase approximation) and in the 
expansion of the self-energy in terms of W (beyond the GW 
approximation). Exchange-correlation effects are described within 
time-dependent density-functional theory from the knowledge of an 
adiabatic {\it nonlocal} XC kernel that goes beyond the local-density 
approximation. 
\end{abstract}


\maketitle

\section{Introduction}

At metal surfaces there exist specific electronic states not present in
the bulk, which can be classified as intrinsic (crystal-induced) surface
states\cite{inglesfield} and image-potential (Rydberg-like)
states.\cite{smith1,imst2} Intrinsic surface states are originated by
the symmetry breaking at the surface, they have their maximum near the
surface, and they are classified as Tamm\cite{tm} and Shockley\cite{sh}
states; in particular, intrinsic Shockley surface states typically occur
in the gap of free-electron-like s,p bands near the Fermi
level.\cite{Gart,plummer} Image-potential states appear as a result of
the self-interaction that an electron near the surface suffers from the
polarization charge it induces at the surface, and they occur in the
vacuum region of metal surfaces with a band gap near the vacuum
level.\cite{imst1}

\begin{figure}
\includegraphics[angle=-90,scale=0.3]{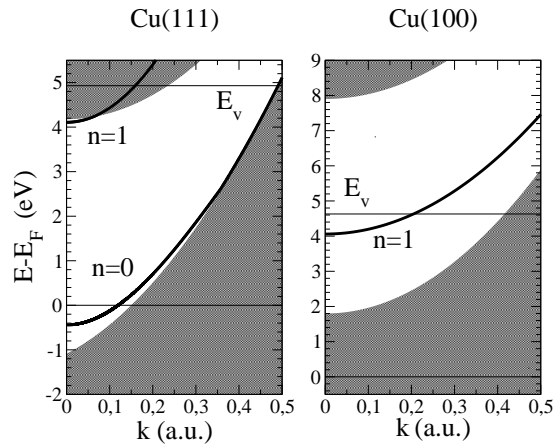}
\caption{The $\Gamma L$ projected bulk band 
structure (shaded areas) of the (111) and (100) surfaces of Cu. The 
solid lines represent Shockley ($n=0$) and image-potential ($n=1$) 
surface-state bands.} 
\label{fig:chulpot}
\end{figure}

Figure~\ref{fig:chulpot} illustrates Shockley and image-potential states in the
gap of the $\Gamma L$ projected band structure of the (100) and (111) surfaces 
of Cu. If an electron or hole is added to the solid at one of these 
states, inelastic coupling of the excited quasiparticle with the 
crystal, which can be experimentally observed through a variety of 
spectroscopies,\cite{exp1,exp2,exp3,fauster,exp4,exp5} may occur 
through electron-electron (e-e) and electron-phonon (e-ph) scattering. 
The decay rate due to the e-ph interaction, which is relatively 
important only in the case of excited Shockley holes near the Fermi 
level, has been investigated recently by using the Eliashberg 
function.\cite{eiguren} Accurate many-body calculations of the decay 
rate due to the e-e interaction were first carried out for image states 
on the (100) and (111) surfaces of Cu.\cite{chulkov1,chulkov2} Since 
then, many-body calculations of the e-e decay have been reported for a 
variety of simple, noble and transition metals.~\cite{sc,s1,fukui,review2}
Nevertheless, existing calculations have been typically performed within the
$G^0W^0$ approximation of many-body 
theory,\cite{hedin,guna,review1,nekovee} with no inclusion of exchange 
and correlation (XC) effects. Exceptions are (i) a calculation of the e-e 
decay rate of image states on the (100) and (111) surfaces of Cu that 
incorporates XC effects in an adiabatic local-density approximation 
(ALDA)\cite{chulkov2} and (ii) an approximate evaluation of the lifetime of 
Shockley states in the noble metals that incorporates the exchange 
contribution to the self-energy.\cite{fukui}

In this paper, we report extensive calculations of the screened 
interaction, the self-energy, and the e-e inelastic lifetime of Shockley 
and image states on the (100) and (111) surfaces of Cu that go beyond 
the $G^{0}W^{0}$ approximation. Short-range XC effects are incorporated both 
in the description of the dynamical screening of the many-electron system 
[we go beyond the random-phase approximation (RPA) in the evaluation of 
the screened interaction W] and in the expansion of the electron 
self-energy in terms of $W$ [we go beyond the $GW$ approximation]. This 
is the $GW\Gamma$ approximation of many-body theory,\cite{mahan,mahan2} 
which treats on the same footing XC effects between pairs of electrons 
within the Fermi sea (screening electrons) and between the excited 
electron and the Fermi sea.

Mahan and Sernelius\cite{mahan} showed that the inclusion, within the 
$GW\Gamma$ approximation, of the same vertex function in the screened 
interaction and the numerator of the self-energy yields results for the 
band-width of a homogeneous electron gas very similar to those obtained 
in the $G^0W^0$ approximation, due to a large cancellation of vertex 
corrections. Large cancellations were also observed to occur in the 
decay rates of image\cite{chulkov2} and bulk\cite{gurtubay} states in 
the noble metals, by incorporating XC effects in the ALDA. In the decay 
of low-energy bulk states below the vacuum level energy transfers 
$\hbar\omega$ are well below the Fermi energy and momentum transfers 
$\hbar q$ are typically smaller than $2\hbar q_F$, $q_F$ being the 
magnitude of the Fermi wave vector, so that one can safely assume that 
both $q$ and $\omega$ are small and XC effects can, therefore, be 
incorporated in the ALDA. However, in the case of Shockley and image 
states the ALDA might lead to spurious results, due to the presence of 
small local values of the Fermi wave vector in a region where the 
electron density is small. Hence, here we use an adiabatic {\it 
nonlocal} XC kernel that accurately describes XC effects in the limit of 
a homogeneous electron gas of arbitrary density and which has been 
succesful in the description of the XC contribution to the jellium 
surface energy.\cite{pp}

It has been argued in the past that a realistic first-principles description of
the electronic band structure is of key importance in the determination of the
inelastic lifetime of bulk electronic states in the noble
metals.\cite{campillo} The main conclusion drawn in Ref.~\onlinecite{campillo}
was that in the case of the noble metals deviations from electron dynamics in
a free gas of $sp$ electrons mainly originate in the participation of $d$
electrons in the screening of electron-electron interactions. The role of
occupied $d$ bands in the dynamics of excited surface-state electrons and
holes on silver surfaces was later investigated via a polarizable medium
giving rise to additional screening,~\cite{ag} and it was concluded that $d$
electrons do not participate significantly in the screening of the interaction
between surface states (which are located near the surface) and the Fermi gas
of the solid.\cite{notenew}

In order to investigate the dynamics of Shockley and image states on Cu
surfaces, we assume that the density of valence electrons in the solid varies
only along the $z$ axis, which is taken to be normal to the surface. Hence, 
our calculations start by solving the single-particle time-independent 
Schr\"odinger equation of electrons moving in a physically motivated
one-dimensional (1D) model potential that is known to correcly reproduce the
behaviour of $sp$ valence states and accurately describes, in particular, the
projected bulk 
band gap and the binding energy of the Shockley and the first image 
state.\cite{chulkov0} The eigenfunctions and eigenvalues of such an 
effective single-particle hamiltonian are then used to compute the 
screened interaction, the $GW\Gamma$ self-energy, and the e-e decay 
rates of Shockley and image states. For comparison, we also compute 
$G^0W^0$, $G^0W$, and $GW^0\Gamma$ decay rates, with no inclusion of XC 
effects, with inclusion of XC effects beyond the RPA in the screened 
interaction $W$ alone, and with inclusion of XC effects beyond the $G^0W^0$ in 
the expansion of the electron self-energy in terms of the RPA screened 
interaction $W^0$, respectively. Our results indicate that (i) although 
the use of the ALDA leads to spurious results for the screened 
interaction, a more realistic adiabatic {\it nonlocal} description of XC 
effects yields inelastic lifetimes of Shockley and image states that 
esentially coincide with those obtained in the ALDA, and (ii) the 
overall effect of short-range XC is small and $GW\Gamma$ linewidths are 
close to their $G^0W^0$ counterparts, as occurs in the case of 
low-energy bulk states.\cite{gurtubay}

The paper is organized as follows. Explicit expressions for the e-e 
decay rate of surface-state electrons and holes at solid surfaces are 
derived in Sec. II, in the $GW\Gamma$ approximation of many-body theory. 
The results of numerical calculations of the screened interaction, the 
self-energy, and the decay rate of Shockley and image states on the 
(100) and (111) surfaces of Cu are presented in Sec. III. The summary 
and conclusions are given in Sec. IV. Unless stated otherwise, atomic 
units (a.u.) are used throughout, i.e., $e^2=\hbar=m_e=1$.
  
\section{Theory}

Let us consider an arbitrary many-electron system of density $n_0({\bf 
r})$. In the framework of many-body theory, the decay rate (or 
reciprocal lifetime) of a quasiparticle (electron or hole) that has been 
added in the single-particle state $\phi_i({\bf r})$ of energy 
$\varepsilon_i$ is obtained as the projection of the imaginary part of 
the self-energy $\Sigma({\bf r},{\bf r}';\varepsilon_i)$ over the 
quasiparticle-state itself\cite{review1}
\begin{equation}\label{eq1}
\tau_i^{-1}=\mp 2\int d{\bf r}\int d{\bf r}'\phi^{*}_i({\bf r})
{\rm Im}\Sigma({\bf r},{\bf r}';\varepsilon_i)\phi_i({\bf r}'),
\end{equation}
where the $\mp$ sign in front of the integral should be taken to be
minus or plus depending on whether the quasiparticle is an electron
($\varepsilon_i\ge\varepsilon_F$) or a hole
($\varepsilon_i\le\varepsilon_F$), respectively, $\varepsilon_F$ being
the Fermi energy.

To lowest order in a series-expansion of the self-energy in terms of 
the frequency-dependent screened interaction $W({\bf r},{\bf
r}';\omega)$, the self-energy is obtained by integrating the product of
the interacting Green function $G({\bf r},{\bf
r}';\varepsilon_i-\omega)$ and the screened interaction $W({\bf r},{\bf
r}';\omega)$, and is therefore called the $GW$ self-energy. If one
further replaces the interacting Green function by its noninteracting
counterpart $G^0({\bf r},{\bf r}';\varepsilon_i-\omega)$, one finds the
$G^0W$ self-energy and from Eq.~(\ref{eq1}) the following expression 
for the $G^0W$ reciprocal lifetime:
\begin{eqnarray}\label{eq:tau}
\tau_i^{-1}&=&\mp 2\,\sum_f\int{\rm d}{{\bf r}}\int{\rm d}{{\bf
r}'}\,\phi_{i}^*({\bf r})\,\phi_{f}^*({\bf
r}')\cr\cr&\times&{\rm Im}\,W({\bf r},{\bf
r}';|\varepsilon_i-\varepsilon_f|)\,\phi_{i}({\bf r}')\,\phi_{f}({\bf
r}),
\end{eqnarray}
where the sum is extended over a complete set of single-particle states
$\phi_f({\bf r})$ of energy $\varepsilon_f$
($\varepsilon_F\leq\varepsilon_f\leq\varepsilon_i$ or
$\varepsilon_i\leq\varepsilon_f\leq \varepsilon_F$).
Equation~(\ref{eq:tau}) exactly coincides with the result one would
obtain from the lowest-order probability per unit time for an excited
electron or hole in an initial state $\phi_i({\bf r})$ of energy
$\varepsilon_i$ to be scattered into the state $\phi_f({\bf r})$ of
energy $\varepsilon_f$ by exciting a Fermi system of interacting
electrons from its many-particle ground state to some many-particle
excited state.\cite{pi1}

The  interaction $W({\bf r},{\bf r}';\omega)$ entering
Eq.~(\ref{eq:tau}) can be rigurously expressed as follows
\begin{eqnarray}\label{eq:Wrpa}W({\bf r},{\bf
r}';\omega)&=&v({\bf r},{\bf r}')+\int{\rm d}{\bf r}_1\int{\rm d}{\bf
r}_2\,v({\bf r},{\bf r}_1)\cr\cr&\times&\chi({\bf r}_1,{\bf
r}_2;\omega)\,v({\bf r}_2,{\bf r}'),
\end{eqnarray}
$v({\bf r},{\bf r}')$ representing the bare Coulomb interaction and
$\chi({\bf r},{\bf r}';\omega)$ being the time-ordered density-response
function of the many-electron system, which for the positive 
frequencies ($\omega>0$) entering Eq.~(\ref{eq:tau}) coincides with the
retarded density-response function of linear-response theory. In the
framework of time-dependent density-functional theory (TDDFT),\cite{tddft}
the {\it exact} retarded density-response function is obtained by solving the
following integral equation:\cite{petersilka}
\begin{eqnarray}\label{eq:Xalda}
&&\chi({\bf r},{\bf r}';\omega)=\chi^0({\bf r},{\bf 
r}';\omega)+\int{\rm d}{\bf
r}_1\int{\rm d}{\bf r}_2\,\chi^0({\bf r},{\bf r}_1;\omega)\cr
\cr&&\times\left\{v({\bf r}_1,{\bf r}_2)+f^{xc}[n_0]({\bf r}_1,{\bf
r}_2;\omega)\right\}\chi({\bf r}_2, {\bf r}';\omega),
\end{eqnarray}
where $\chi^0({\bf r},{\bf r}';\omega)$ denotes the density-response
function of noninteracting Kohn-Sham electrons, i.e., independent
electrons moving in the effective Kohn-Sham potential of density-functional
theory (DFT). The frequency-dependent XC kernel $f^{xc}[n_0]({\bf r},{\bf r}'
\omega)$ is
the functional derivative of the frequency-dependent XC potential
$V_{xc}[n]({\bf r},\omega)$ of TDDFT, to be evaluated at $n_0({\bf 
r})$. In the RPA, $f^{xc}[n_0]({\bf r},{\bf r}';\omega)$ is set equal 
to zero and Eq.~(\ref{eq:tau}) yields the so-called $G^0W^0$ (or
$G^0W$-RPA) reciprocal lifetime.

The xc kernel $f^{xc}[n_0]({\bf r},{\bf r}';\omega)$, which is absent 
in the RPA, accounts for the presence of an XC hole associated to 
all screening electrons in the Fermi sea. Hence, one might be tempted to
conclude that the full $G^0W$ approximation [with the formally exact
screened interaction $W$ of Eq.~(\ref{eq:Wrpa})] should be a better
approximation than its $G^0W^0$ counterpart [with the screened
interaction $W$ evaluated in the RPA]. However, the XC hole associated
to the excited hot electron is still absent in the $G^0W$ approximation.
Therefore, if one goes beyond RPA in the description of
$W$, one should also go beyond the $G^0W$ approximation in the expansion
of the electron self-energy in powers of $W$. By including XC effects
both beyond RPA in the description of $W$ and beyond $G^0W$ in the
description of the self-energy,\cite{mahan,mahan2} the so-called
$GW\Gamma$ approximation yields a lifetime broadening that is of the
$G^0W$ form [see Eq.~(\ref{eq:tau})], but with the actual screened
interaction $W({\bf r},{\bf r}';\omega)$ of Eq.~(\ref{eq:Wrpa}) replaced by a
new effective screened interaction
\begin{eqnarray}\label{eq:Walda}
&&\tilde W({\bf r},{\bf r}';\omega)
=v({\bf r},{\bf r}')+\int{\rm d}{\bf r}_1\int{\rm
d}{\bf r}_2\,\left\{v({\bf r},{\bf
r}_1)\right.\cr\cr&&+\left.f^{xc}[n_0]({\bf r},{\bf
r}_1;\omega)\right\}\,\chi({\bf r}_1,{\bf r}_2;\omega)\,v({\bf r}_2,{\bf
r}'),
\end{eqnarray}
which includes all powers in $W$ beyond the $G^0W$ approximation.

\subsection{Bounded electron gas}

In the case of a bounded electron gas that is translationally invariant
in two directions, such as the jellium surface or the physically
motivated model surface described above, the single-particle states
entering Eq.~(\ref{eq:tau}) are of the form
\begin{equation}\label{bounded1}
\phi_{{\bf k},i}({\bf r})=\phi_i(z){\rm e}^{i{\bf k}\cdot{\bf r}_\parallel}
\end{equation}
with energies
\begin{equation}\label{bounded2}
\varepsilon_{{\bf k},i}=\varepsilon_i+k^2/2m_i,
\end{equation}
${\bf k}$ being a wave vector parallel to the surface and $m_i$ denoting
the effective mass in the plane of the surface.\cite{note1}

Introducing Eqs.~(\ref{bounded1}) and (\ref{bounded2}) into Eq.~(\ref{eq1}),
one finds the following expression for the reciprocal lifetime of a
quasiparticle (electron or hole) that has been added in the single-particle
state 
$\phi_{{\bf k},i}({\bf r})$ of energy $\varepsilon_{{\bf k},i}$:
\begin{equation}\label{eqn}
\tau_{{\bf k},i}^{-1}=\mp 2\int dz\int dz'\phi^{*}_i(z)
{\rm Im}\Sigma(z,z';{\bf k},\varepsilon_{{\bf k},i})\phi_i(z'),
\end{equation}
where $\Sigma(z,z';{\bf k},\varepsilon_{{\bf k},i})$ represents the
two-dimensional (2D) Fourier transform of the self-energy
$\Sigma({\bf r},{\bf r}';\varepsilon_{{\bf k},i})$.

\subsubsection{$G^0W$ approximation}

Using the single-particle wave functions and energies of
Eqs.~(\ref{bounded1}) and (\ref{bounded2}), the $G^0W$ reciprocal
lifetime of Eq.~(\ref{eq:tau}) yields
\begin{eqnarray}\label{eq:tau2}
\tau_{{\bf k},i}^{-1}&=&\mp 2\,\sum_f\int{d{\bf q}\over(2\pi)^2}\int{\rm
d}z\int{\rm d}z'\,\phi_{i}^*(z)\,\phi_{f}^*(z')\cr\cr&\times&{\rm 
Im}\,W(z,z';{\bf q},\omega)\,\phi_{i}(z')\,\phi_{f}(z),
\end{eqnarray}  
where $\omega=|(\varepsilon_i+k^2/2m_i)-(\varepsilon_f+q^2/2m_f)|$, 
${\bf k}$ and ${\bf q}$ represent wave vectors parallel to the surface, 
and $W(z,z';{\bf k},\omega)$ denotes the 2D Fourier 
transform of the screened interaction $W({\bf r},{\bf r}';\omega)$ of 
Eq.~(\ref{eq:Wrpa}), i.e.,
\begin{eqnarray}\label{eq:Wrpa2}
W(z,z';{\bf k},\omega)&=&v(z,z';{\bf k})+\int{\rm d}z_1\int{\rm 
d}z_2\,v(z,z_1;{\bf k})\cr\cr&\times&\chi(z_1,z_2;{\bf 
k},\omega)\,v(z_2,z';{\bf k}),
\end{eqnarray}
$v(z,z';{\bf k})$ and $\chi(z,z';{\bf k},\omega)$ being 2D Fourier 
transforms of the bare Coulomb interaction and the density-response 
function of Eq.~(\ref{eq:Xalda}), respectively. 

In the $G^0W^0$ (or $G^0W$-RPA) approximation, the reciprocal lifetime 
is also given by Eqs.~(\ref{eq:tau2}) and (\ref{eq:Wrpa2}), but with the XC 
kernel $f^{xc}[n_0]({\bf r},{\bf r}';\omega)$ entering 
Eq.~(\ref{eq:Xalda}) set equal to zero.

\subsubsection{$GW\Gamma$ approximation}
\label{gamma}

Using the single-particle wave functions and energies of 
Eqs.~(\ref{bounded1}) and (\ref{bounded2}), the $GW\Gamma$ reciprocal 
lifetime is also found to be given by Eq.~(\ref{eq:tau2}), but with 
$W(z,z';{\bf k},\omega)$ replaced by the 2D Fourier transform of the
effective screened interaction $\tilde W({\bf r},{\bf r}';\omega)$ of
Eq.~(\ref{eq:Walda}), 
i.e.:
\begin{eqnarray}\label{eq:tau3}
\tau_{{\bf k},i}^{-1}&=&\mp 2\,\sum_f\int{d{\bf q}\over(2\pi)^2}\int{\rm 
d}z\int{\rm d}z'\,\phi_{i}^*(z)\,\phi_{f}^*(z')\cr\cr&\times&{\rm 
Im}\,\tilde W(z,z';{\bf q},\omega)\,\phi_{i}(z')\,\phi_{f}(z),
\end{eqnarray}  
where
\begin{eqnarray}\label{eq:Walda3}
&&\tilde W(z,z';{\bf k},\omega)
=v(z,z';{\bf k})+\int{\rm d}z_1\int{\rm
d}z_2\,\left\{v(z,z_1;{\bf 
k})\right.\cr\cr&&+\left.f^{xc}[n_0](z,z_1;{\bf 
k},\omega)\right\}\,\chi(z_1,z_2;{\bf k},\omega)\,v(z_2,z';{\bf k}),
\end{eqnarray}
$f^{xc}[n_0](z,z_1;{\bf k},\omega)$ being the 2D Fourier transform of 
the XC kernel $f^{xc}[n_0]({\bf r},{\bf r}';\omega)$.

In the $GW^0\Gamma$ approximation, the reciprocal lifetime is also given by 
Eqs.~(\ref{eq:tau3}) and (\ref{eq:Walda3}), thereby with full inclusion 
of the XC kernel entering Eq. (\ref{eq:Walda3}), but with the XC kernel 
entering Eq.~(\ref{eq:Xalda}) set equal to zero.

Hence, we note that both $G^0W$ and $GW\Gamma$ reciprocal lifetimes 
[Eqs.~(\ref{eq:tau2}) and (\ref{eq:tau3})] can be calculated from the 
knowledge of two basic ingredients: (i) single-particle wave functions and 
energies of the form of Eqs.~(\ref{bounded1}) and (\ref{bounded2}), 
which are also basic quantities in the evaluation of the noninteracting 
density-response function $\chi^0(z,z';{\bf k},\omega)$, and (ii) the XC 
kernel $f^{xc}[n_0](z,z';{\bf k},\omega)$. 

\subsection{Single-particle wave functions and energies}
\label{wave}

For the description of the noninteracting density-response function
$\chi^0(z,z';{\bf k},\omega)$ [and, therefore, the screened interaction
$W(z,z';{\bf k},\omega)$ and the effective screened interaction
$\tilde W(z,z';{\bf k},\omega)$ of Eqs.~(\ref{eq:Wrpa2}) and (\ref{eq:Walda3}),
respectively] single-particle wave functions and energies can safely be taken
to be the eigenvalues and eigenfunctions of a {\it jellium} self-consistent
Kohn-Sham hamiltonian.~\cite{chulkov2} Nevertheless, the actual band structure
of
$sp$ electrons near the surface of noble metals calls for a more realistic
description of the single-particle wave functions [$\phi_i(z)$ and
$\phi_f(z)$] and energies [$\varepsilon_i$ and $\varepsilon_f$]
entering Eqs.~(\ref{eq:tau2}) and (\ref{eq:tau3}).

Hence, in the calculations presented in this paper all the single-particle wave
functions and energies (those entering Eqs.~(\ref{eq:tau2}) and
(\ref{eq:tau3}) and also those involved in the evaluation of the
noninteracting density-response function) are taken to be the eigenfunctions
and eigenvalues of a physically motivated 1D model hamiltonian that accurately
reproduces the projected 
band gap and the binding energy of the Shockley and the first image 
state.\cite{chulkov0}

\subsection{The XC kernel $f^{xc}[n_0](z,z';{\bf k},\omega)$} 

In order to investigate the impact of strong variations of the 
electron density induced near the surface, and because the excitation energies 
of interest are typically small (particularly in the case of Shockley 
holes), we consider the following {\it adiabatic} ($\omega=0$) approximations 
of the XC kernel $f^{xc}[n_0](z,z';{\bf k},\omega)$:\cite{pp}

\subsubsection{Adiabatic local-density approximation (ALDA)}

If one assumes that dynamic electron-density fluctuations are slowly 
varying in all directions, the XC kernel $f^{xc}[n_0](z,z';{\bf 
k},\omega)$ is easily found to be given by the following 
expression:\cite{liebsch}
\begin{equation}\label{alda}
f^{xc}[n_0](z,z';{\bf k},\omega)=\bar 
f^{xc}(n_0(z);k^{3D}=0,\omega=0)\,\delta(z-z').
\end{equation}
Here, $\bar f^{xc}(n_0(z);k^{3D},\omega)$ is the 3D Fourier transform of 
the XC kernel of a homogeneous electron gas of density $n_0(z)$, which 
in the limit as $k^{3D}\to 0$ and $\omega\to 0$ is known to be the second 
derivative of the XC energy $\varepsilon_{xc}(n)$ per particle of a homogeneous
electron gas, to be evaluated at the local density $n_0(z)$. We use the
Perdew-Wang parametrization~\cite{pw} of the difussion Monte Carlo (DMC) XC
energy
$\varepsilon_{xc}$ reported by Ceperley and Alder~\cite{ca}.

\subsubsection{Refined ALDA}

A more accurate description of short-range XC effects can be carried out 
by replacing the {\it local} XC kernel $\bar 
f^{xc}(n_0(z);k^{3D}=0,\omega=0)$ entering Eq.~(\ref{alda}) by a more 
accurate still adiabatic but momentum-dependent XC kernel $\bar 
f^{xc}(n_0(z),k^{3D}=k,\omega=0)$ (thus only assuming that the dynamic 
density fluctuation is slowly varying in the direction perpendicular to 
the surface), i.e,
\begin{equation}\label{refined}
f^{xc}[n_0](z,z';{\bf k},\omega)=\bar 
f^{xc}(n_0(z);k^{3D}=k,\omega=0)\,\delta(z-z').
\end{equation}

Here we exploit the accurate DMC calculations reported by Moroni {\it et
al.}\cite{moroni} for the static ($\omega=0$) $k^{3D}$-dependent
{\it nonlocal} XC kernel $\bar f^{xc}$ of a homogeneous electron gas. A
parametrization of this data satisfying the well-known small- and
large-wavelength asymptotic behaviour was carried out by Corradini
{\it et al.} (CDOP)~\cite{corradini}.

\subsubsection{Adiabatic {\it nonlocal} approximation (ANLDA)}
\label{anla}

Here we still neglect the frequency dependence of the XC kernel (adiabatic
approximation), but now we make no assumption on the variation of the
dynamic density fluctuation and assume that the {\it unperturbed} density
variation $\left[n_0(z)-n_0(z')\right]$ is small within the short range of 
$f^{xc}[n_0](z,z';{\bf k},\omega)$. This allows to write 
\begin{equation}\label{one}
f^{xc}[n_0](z,z';{\bf k},\omega)=\bar 
f^{xc}(\left[n_0(z)+n_0(z')\right]/2;z,z';k,\omega=0),
\end{equation}  
where $\bar f^{xc}(n;z,z';k,\omega)$ represents the 2D Fourier transform of the
XC kernel $\bar f^{xc}(n;k,\omega)$ of a homogeneous electron gas of density
$n$. An explicit expression
for the 2D Fourier transform of the CDOP parametrization of
$\bar f^{xc}(n;k,\omega=0)$ was reported in Ref.~\onlinecite{pp}:
\begin{widetext}
\begin{equation}\label{corradini}
\bar f^{xc}(n;z,z';k)=-\frac{4\pi e^2 C}{k_F^2}\delta(\tilde 
z)-\frac{2\pi
e^2 B}{\sqrt{gk_F^2+k^2}}\,{\rm e}^{-\sqrt{gk_F^2+k^2}|\tilde z|}-
\frac{2\alpha\sqrt{\pi/\beta}e^2}{k_F^3}\left[\frac{2\beta-k_F^2\tilde
z^2}{4\beta^2}k_F^2+k^2\right] {\rm e}^{-\beta\left[k_F^2\tilde
z^2/4\beta^2
+k^2/k_F^2\right]},
\end{equation}
\end{widetext}
where $C$, $B$, $g$, $\alpha$, and $\beta$ are dimensionless functions
of the electron density (see Ref.~\onlinecite{corradini}),
$n=k_F^3/3\pi^2$, and $\tilde z=z-z'$.

\section{Results and discussion}
\label{sec:R}

On the (111) surface of Cu, the $n=0$ Shockley state at the center of
the surface Brillouin zone ($k=0$) lies just below the
Fermi level, with $\varepsilon_i-\varepsilon_F=-0.445\,{\rm eV}$.
Binding energies of the $n=1$ image state on the (111) and (100)
surfaces of Cu (measured with respect to the vacuum level) are 0.83 and
0.57 meV, respectively. Effective masses of the $n=1$ image state on
Cu(111) and Cu(100) are close to the free-electron mass
($m_i=1$),\cite{fauster} while the effective mass of the $n=0$ Shockley
state on Cu(111) is 0.42.\cite{m1,m2} The probability density of the
$n=1$ image states on Cu(111) and Cu(100) have a maximum at 2.3 and 3.8
${\rm\AA}$, respectively, outside the crystal edge ($z=0$), which we
choose to be located half a lattice spacing beyond the last atomic
layer. The $n=0$ Shockley state wave function in Cu(111), however, is
maximum at the crystal edge.

\subsection{Screened interaction}

We have carried out calculations of the imaginary part of the screened
interaction $W(z,z';{\bf k},\omega)$ and the effective screened interaction
$\tilde W(z,z';{\bf k},\omega)$ of thin slabs. In order to ensure that our
slab calculations are a faithful representation of the actual screened
interaction of a semiinfinite system, we have used films up to 50 layers of
atoms and 80 interlayer-spacing vacuum intervals, as in the $G^0W^0$
($G^0W$-RPA) calculations reported in Refs.~\onlinecite{chulkov1} and
\onlinecite{chulkov2}.

\begin{figure}
\includegraphics[angle=-90,scale=0.3]{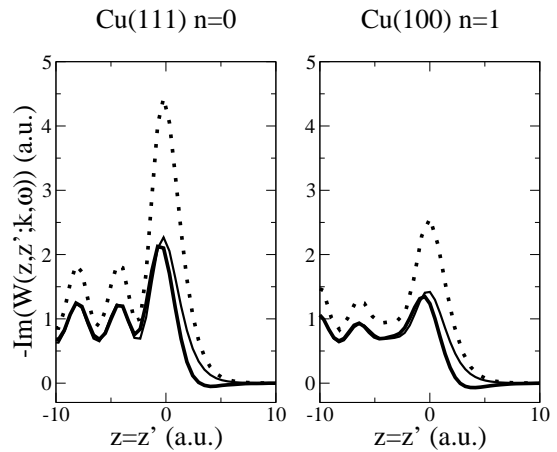}
\caption{Imaginary part of the screened interaction $W(z,z';{\bf k},\omega)$
and the effective screened interaction $\tilde W(z,z';{\bf k},\omega)$, as a
function of $z=z'$ and for fixed values of $k$ and $\omega$ ($k=0.5\
{\rm\AA}^{-1}$ and
$\omega=0.5\,{\rm eV}$), in the vicinity of the (100) and (111) surfaces of Cu.
ALDA calculations of ${\rm Im}\left[\tilde W(z,z';{\bf k},\omega)\right]$ are
represented by thick solid lines. RPA and ALDA calculations of ${\rm
Im}\left[W(z,z';{\bf k},\omega)\right]$ are represented by thin solid  and
dotted lines, respectively.}
\label{fig:imw_chul_jel}
\end{figure}

The impact of XC effects on the imaginary part of the effective screened 
interaction in the vicinity of the (100) and (111) surfaces of Cu is
illustrated in Fig.~\ref{fig:imw_chul_jel}, where ALDA calculations of
${\rm Im}[\tilde W(z,z';{\bf k},\omega)]$ (with full
inclusion of XC effects) are compared to calculations of
${\rm Im}[W(z,z';{\bf k},\omega)]$ with (ALDA) and without (RPA) XC effects.
Exchange-correlation effects included in the effective screened
interaction have two sources, as discussed in Section~\ref{gamma}.
First, there is the reduction of the screening due to the presence of an
XC hole associated to all electrons in the Fermi sea [see
Eq.~(\ref{eq:Xalda})], which is included in the calculations represented in
Fig.~\ref{fig:imw_chul_jel} by thick solid lines and {\it also} in the
calculations represented by dotted lines. Secondly, there is the reduction of
the effective screened interaction itself due to the XC hole associated to
each
electron [see Eq.~(\ref{eq:Walda3})], which is {\it only} included in the
calculations represented in Fig.~\ref{fig:imw_chul_jel} by thick solid lines.
These contributions have opposite signs and it is the latter which dominates.

\begin{figure}
\includegraphics[angle=-90,scale=0.3]{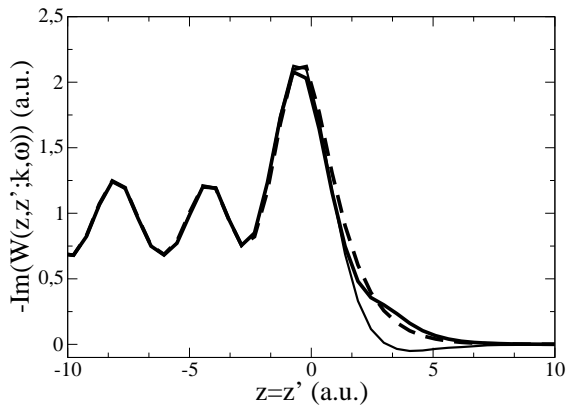}
\caption{Imginary part of the effective screened interaction
$\tilde W(z,z;{\bf k},\omega)$, as a function of $z=z'$ and for fixed values of
$q$ and $\omega$ ($q=0.5\,{\rm\AA}^{-1}$ and $\omega=0.5\,{\rm eV}$), in the
vicinity of the (111) surface of Cu. ALDA, refined ALDA, and ANLDA
calculations are represented by thin solid, dashed, and thick solid lines,
respectively.}
\label{fig:figure4}
\end{figure}

Existing $GW\Gamma$ calculations of the lifetime broadening of image 
states on Cu(100) and Cu(111) were performed with the ALDA XC kernel
that we have used in the calculations represented in
Fig.~\ref{fig:imw_chul_jel}. The error introduced by the use of this {\it
local} kernel is small in the
interior of the solid, as the wave vectors involved are small ($k<k_F$).
However, Fig.~\ref{fig:imw_chul_jel} shows that the ALDA leads to spurious
(negative) results for ${\rm Im}[\tilde W(z,z';{\bf k},\omega)]$ near the
surface, which is due to the presence of small local values of the Fermi wave
vector ($k_F^{local}<k$) in a region where the electron density is small.
Hence, we have carried out refined ALDA and adiabatic {\it nonlocal} (ANLDA)
calculations of ${\rm Im}[\tilde W(z,z';{\bf k},\omega)]$ (both with full
inclusion
of XC effects), which have been plotted in Fig.~\ref{fig:figure4}. This figure
clearly shows that the impact of {\it nonlocality} on the effective screened
interaction is large near the surface, bringing spurious ALDA calculations
(thin solid lines) to a more realistic behaviour near the surface (thick solid
lines). The refined ALDA scheme partially overcomes the failure of the ALDA,
but a full description of the nonlocality of XC effects near the surface might
be needed for a realistic description of the absorption power of solid
surfaces.

\subsection{Self-energy}

\begin{figure}
\includegraphics[angle=-90,scale=0.35]{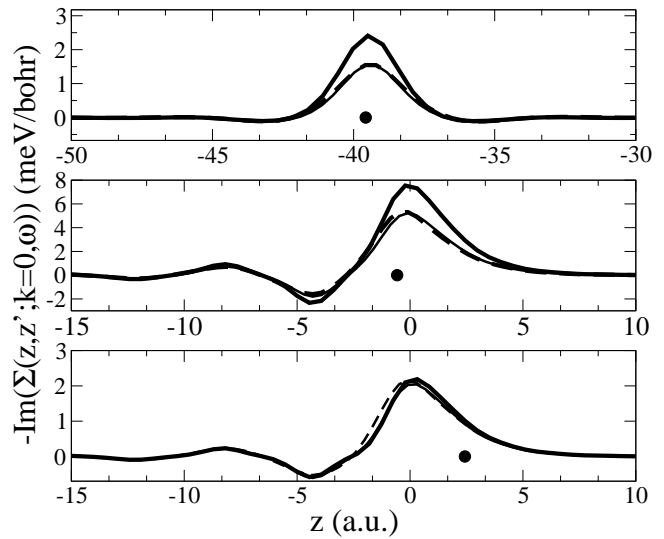}
\caption{$G^0W^0$ ($G^0W$-RPA), $G^0W$, and $GW\Gamma$ calculations of the
imaginary part of the $n=0$ surface-state self-energy 
$\Sigma(z,z';{\bf k}=0,\varepsilon_{\bf k})$, versus $z$, in the vicinity of
the (111) surface of Cu. The solid circle represents the value of $z'$ in ech
case. $GW\Gamma$ calculations (as obtained with the use of our ANLDA XC
kernel) are represented by dashed lines. $G^0W$ (also using our ANLDA XC
kernel) and $G^0W^0$ calculations are represented by thin and thick solid
lines, respectively. ALDA calculations, which nearly coincide with ANLDA
calculations, are not plotted in this figure.}
\label{multiself}
\end{figure}

Figure~\ref{multiself} exhibits $G^0W^0$ ($G^0W$-RPA), $G^0W$, and
$GW\Gamma$ calculations of the imaginary part of the $n=0$ surface-state
self-energy $\Sigma(z,z';{\bf k}=0,\varepsilon_{\bf k})$, versus $z$, in the
vicinity of the (111) surface of Cu, with use (in the case of the $G^0W$ and
$GW\Gamma$ approximations) of the adiabatic {\it nonlocal} XC kernel (ANLDA)
described in section~\ref{anla}. This figure shows that as occurs in the case
of the screened interaction XC effects partially compensate each other,
leading to an overall effect of no more than $5\%$. For comparison, we have
also used (in the case of the $G^0W$ and $GW\Gamma$ approximations) the ALDA
and refined ALDA kernels described in section~\ref{anla}, and we have found 
that although the use of these local or semilocal kernels leads to spurious
results for the screened interaction, our more realistic ANLDA kernel yields
self-energies that esentially coincide with those obtained in the ALDA.   

\subsection{Reciprocal lifetime}

\begin{figure}
\includegraphics[angle=-90,scale=0.3]{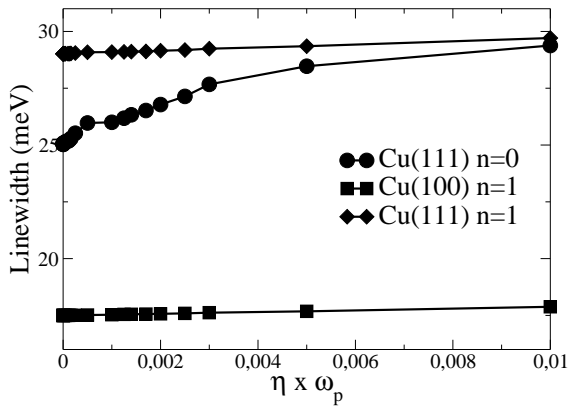}
\caption{$G^0W^0$ reciprocal lifetimes of Shockley and image states on the
(100) and (111) surfaces of Cu, as a function of the parameter $\eta$ that
accounts for the imaginary part of the complex frequencies entering the
evaluation of the noninteracting density-response function $\chi^0(z,z';{\bf
k},\omega)$.}
\label{fig:fig2}
\end{figure}

Now we focus on the evaluation of the decay rate (reciprocal lifetime)
of surface-state electrons (and holes) at the $n=1$ (and $n=0$)
surface-state band edge (${\bf k}=0$) of the (111) and (100) surfaces of
Cu. Calculations of the noninteracting density-response function
$\chi^0(z,z';{\bf k},\omega)$ (and, therefore, the reciprocal lifetime) require
the introduction of complex frequencies of the form $\omega+{\rm i}\eta$,
$\eta$ being a positive infinitesimal. Hence, in order to ensure that our
numerical calculations yield a converged value of the reciprocal lifetime, we
have calculated $\tau^{-1}$ as a function of the parameter $\eta$.
Fig.~\ref{fig:fig2} represents the results we have obtained for the $G^0W^0$
reciprocal lifetimes of Shockley and image states on the (100) and (111)
surfaces of Cu, showing that converged results are obtained for a sufficiently
small value of $\eta$.

\begin{table}
\caption{$G^{0}W^{0}$, $G^{0}W$, and $GW\Gamma$ reciprocal lifetimes, in
linewidth units (meV), of an excited surface-state electron (hole) at the
$n=1$ ($n=0$) surface-state band edge (${\bf k}=0$) of the (111) and (100)
surfaces of Cu. In the case of the $G^{0}W$ and $GW\Gamma$ reciprocal
lifetimes, both ALDA and ANLDA exchange-correlation kernels have been
considered.}\label{tableD0}
\begin{ruledtabular}
\begin{tabular}{lcccccccc}
Surface&$n$& XC kernel & $G^{0}W^{0}$ & $G^{0}W$ & $GW\Gamma$\\
\hline
Cu(100)&1&    &17.5&    &    \\
       &1&ALDA&    &24&17\\
       &1&ANLDA&   &24.5&17\\
Cu(111)&0&    &25&    &    \\
       &0&&    &30&24.5\\
       &0&ANLDA&   &30.5&24.5\\
Cu(111)&1&    &29&    &    \\ 
       &1&ALDA&    &42.8&28.5\\  
       &1&ANLDA&   &43&28\\   
\end{tabular}
\end{ruledtabular}
\end{table}

Converged calculations of the reciprocal lifetimes of Shockley and image states
on the (100) and (111) surfaces of Cu are exhibited in Table~\ref{tableD0}.
This table shows: (i) $G^0W^0$ results, which reproduce previous
calculations~\cite{chulkov1,chulkov2}, (ii) ALDA $GW\Gamma$ results, which in
the case of the $n=1$ image state on Cu(111) and Cu(100) reproduce the
calculations reported in Ref.~\onlinecite{chulkov2} (ALDA $GW\Gamma$
calculations of the reciprocal lifetime of $n=0$ Shockley states had
{\it not} been reported before), and (iii) ANLDA $GW\Gamma$ calculations, never
reported before; for comparison, $G^0W$ reciprocal lifetimes are also shown in
this table, with use of both the ALDA and the adiabatic {\it nonlocal} kernel
ANLDA described in section~\ref{anla}. Differences between our $G^0W^0$
reciprocal lifetime of the $n=0$ Shockley state in Cu(111)
($\tau^{-1}=25\,{\rm meV}$) and those reported before\cite{gpc}
($\tau^{-1}=19\,{\rm meV}$) are simply due to the fact that in our present
model we are not accounting for the change of the $z$-dependent surface-state
wave functions $\phi_i(z)$ and $\phi_f(z)$ along the surface-state dispersion
curve.

As in the case of the self-energy, the results shown in Table~\ref{tableD0}
show that (i) a realistic adiabatic {\it nonlocal} description of XC effects
yields reciprocal lifetimes of Shockley and image states that esentially
coincide with those obtained in the ALDA, and (ii) the overall effect of
short-range XC is small  and $GW\Gamma$ reciprocal lifetimes are close to
their $G^0W^0$ counterparts.   

\section{Summary and conclusions}
\label{sec:C}

We have carried out extensive calculations of the self-energy and lifetime of
Shockley and image states on the (100) and (111) surfaces of Cu, in the
framework of the $GW\Gamma$ approximation of many-body theory. This
approximation treats on the same footing XC effects between pairs of electrons
within the Fermi sea  (screening electrons) and between the excited
surface-state electron (or hole) and the Fermi sea. We have included XC
effects within TDDFT from the knowledge of an adiabatic {\it nonlocal} XC
kernel that goes beyond the local-density approximation, and we have found
that these XC contributions (in the screened interaction $W$ and in the
expansion of the self-energy in terms of $W$) have opposite signs and it is
the latter which dominates, leading to $GW\Gamma$ reciprocal lifetimes that
are only slightly lower than their $G^0W^0$ counterparts.

\acknowledgments  
The authors acknowledge partial support by the UPV/EHU, the Basque
Unibertsitate eta Ikerketa Saila, the Spanish Ministerio de Educaci\'on y
Ciencia (Grant No. CSD2006-53), and the EC 6th framework Network of Excellence
NANOQUANTA.


\begin{thebibliography}{99}
\bibitem{notenew} $d$ electrons do play an important role in the case of image
states (but not in the case of Shockley states) on silver surfaces, but this
is simply a consequence of the reduction (in the presence of $d$ electrons) of
the surface-plasmon energy that allows to open a new decay channel. No
surface-plasmon decay channel is opened in the case of the Cu surfaces, since
even in the presence of $d$ electrons the Cu surface-plasmon energy is too
large for this decay channel to be available.
\bibitem{note1} In the case of the jellium surface, the electron motion
is free in the plane of the surface and the effective mass coincides,
therefore, with that of free electrons. In the case of the model
surface, the effective mass is taken from the actual energy dispersion
of the projected band structure.
\end{thebibliography}
\end{document}